\documentclass[preprint,showpacs,preprintnumbers,amsmath,amssymb]{revtex4}
\usepackage{amsmath}
\usepackage{graphicx}
\usepackage{bm}
\newcommand{\be}{\begin{equation}}
\newcommand{\ee}{\end{equation}}
\newcommand{\ben}{\begin{eqnarray}}
\newcommand{\een}{\end{eqnarray}}

\begin{document}
\title{$H(z)$ diagnostics on the nature of dark energy}
\author{Sergio del Campo\footnote{E-mail address:
sdelcamp@ucv.cl}} \affiliation{Instituto de F\'{\i}sica,
Pontificia Universidad Cat\'{o}lica de Valpara\'{\i}so, Avenida
Brasil 2950, Casilla 4059, Valpara\'{\i}so, Chile}
\author{Ram\'{o}n Herrera\footnote{E-mail address: ramon.herrera@ucv.cl}}
\affiliation{Instituto de F\'{\i}sica, Pontificia Universidad
Cat\'{o}lica de Valpara\'{\i}so, Avenida Brasil 2950, Casilla
4059, Valpara\'{\i}so, Chile}
\author{Diego Pav\'{o}n\footnote{E-mail address: diego.pavon@uab.es}}
\affiliation{Departamento de F\'{\i}sica, Facultad de Ciencias,
Universidad Aut\'{o}noma de Barcelona, 08193 Bellaterra
(Barcelona), Spain}
\begin{abstract}
The two dominant components of the cosmic budget today,
pressureles matter and dark energy, may or may not be interacting
with each other. Currently, both possibilities appear compatible
with observational data. We propose several criteria based on the
history of the Hubble factor that can help discern whether they
are interacting and whether dark energy is phantom or quintessence
in nature.
\end{abstract}
\maketitle

\section{Introduction}
Recently, Sahni {\it et al.}  used the luminosity distance $d_{L}
= c (1+z) \int_{0}^{z}\, dz'/H(z')$, valid in spatially flat
Friedmann-Robertson-Walker (FRW) universes, to propose a
diagnostic to tell the cosmological constant  from evolving dark
energy fields \cite{sahni1}. For universes dominated by cold
matter (CM), subscript $m$, and dark energy (DE), subscript $d$,
with constant equation of state, $w = p_{d}/\rho_{d} < -1/3$,
Friedmann's equation takes the form
 \begin{equation}
E^{2}(x) = \Omega_{m0}\, x^{3} \, + \, \Omega_{d0}\, x^{3(1+w)} \,
, \label{friedmann1}
\end{equation}
where $x := 1+z = a_{0}/a$, and $E(x):= H(x)/H_{0}$ is the
normalized Hubble function. In writing this expression it was
implicitly assumed that cold dark matter (CDM) and DE evolved
separately. As usual, $\Omega_{i0}:= \kappa^{2}\, \rho_{i0}/(3
H_{0}^{2})$ with $i = m, d$ stand for the energy densities of the
components in units of the critical density, $\kappa^{2}:= 8\pi G
\,$, and a zero subscript denotes the present-day value of the
corresponding quantity.

Sahni {\it et al.} \cite{sahni1} introduced the redshift dependent
function
\begin{equation}
Om(x) := \frac{E^{2}(x)-1}{x^{3} -1} \, ,
\label{om}
\end{equation}
which for spatially flat FRW cosmologies ($\Omega_{m0}+
\Omega_{d0} = 1$) reduces to $Om(x) = \Omega_{m0}\, + \,
(1-\Omega_{m0}) (x^{3(1+w)}\, -1)(x^{3} - 1)^{-1}$. It has the
interesting feature that for the $\Lambda$CDM model ($\rho_{d}=
{\rm constant}$, $w = -1$), it yields $ Om(x) = \Omega_{m0} \, $,
whereas for quintessence fields ($-1 <w <-1/3$) and phantom fields
($w<-1$) it gives  $Om(x)> \Omega_{m0}$ and $Om(x)< \Omega_{m0}$,
respectively.

Since this diagnostic does not depend on the present values of the
density parameters, $\Omega_{m0}$ and $\Omega_{d0}$, it
circumvents the drawbacks brought about by our comparatively poor
knowledge of them. In fact, it is expected to be useful when
accurate data of $H$ at different redshifts become available. This
seems more feasible than to substantially upgrade our knowledge of
$\Omega_{m0}$ and $\Omega_{d0}$ in the near future.

At present, the $\Lambda$CDM model fits the observational data,
within statistical errors, rather well -see, e.g. \cite{serra}.
Nonetheless if future data come to reveal that  $ Om(x) \neq
\Omega_{m0} \, $, a number of DE alternatives should be considered
including scenarios in which the DE interacts non-gravitationally
with other components.

Here, we assume that DE and CDM, subscript $c$, interact with each
other (whence they do not evolve separately) but not with baryonic
matter, subscript $b$, since interactions with baryons are
strongly constrained by experiments \cite{baryonic-constraints}.
This possibility is being actively considered in the literature
-see \cite{actively} and references therein- since it is rather
natural \cite{NPB-polyakov1, NPB-polyakov2} and more general than
otherwise, it substantially alleviates the coincidence problem
(i.e., ``why are the densities of matter and dark energy of the
same order precisely today"), and agrees with observation.

In this paper we propose several criteria, based on the evolution
of the Hubble factor, to help discriminate models in which DE and
CDM interact with each other non-gravitationally from models in
which they do not, and to tell quintessence DE energy models from
phantom models. Clearly, for these criteria to be useful accurate
data of $H(z)$ are needed. While the present data exhibit large
error bars \cite{simon,daly,gazta} the situation may well improve
comparatively soon thanks to upcoming and planned experiments. The
Baryon Oscillation Spectroscopy Survey (BOSS) 5-years project
\cite{schlegel} aims at measuring the absolute cosmic distance
scale and expansion rate with percent-level precision at redshifts
$z < 0.7$ and $z \simeq 2.5$ using the standard rule furnished by
the baryon acoustic oscillations. The forecast precision for
$H(z)$ at $z =0.35$, $0.6$ and $2.5$ is $1.8\%$, $1.7\%$ and
$1.2\%$, respectively. Data at intermediate and larger redshifts
will partly come  by adapting to $ z > 0$ the method proposed by
Hogan \cite{craig1} which makes use of the gravitational radiation
emitted by black hole  binary inspiral events to determine $H_{0}$
with a precision better than $1\%$. The method may be extended up
to $z \sim 1$ redshifts, and even to $z \simeq 10$ by observing
massive black hole binaries since there will likely be a direct
electromagnetic counterpart of the merger \cite{craig2}. An even
more promising method, proposed by  Corasaniti {\it et al.}
\cite{stefano} to constrain cosmological parameters, exploits the
Sandage-Loeb test \cite{loeb}. It consists in measuring the
redshift of distant objects at two separate times, say $t_{0}$ and
$t_{0} + \Delta t_{0}$, at the present epoch. The redshift
variation of the source (subscript $s$) between these two instants
is, at a very good precision, $\Delta z_{s} \simeq
(\dot{a}(t_{0})\, - \, \dot{a}(t_{s}))/a(t_{s})$ -see Eq. (5) in
\cite{stefano}. This expression, re-written as $H(z_{s}) \simeq
(\Delta z_{s}/\Delta t_{0}) \, - \, H_{0} \, (1+z_{s})$, provides
us with the value of the Hubble factor at the time the source
emitted the electromagnetic signal that is currently arriving at
our detectors (either telescopes or radio-telescopes), modulo we
know $H_{0}$ accurately enough. The authors of \cite{stefano}
suggested using the absorption lines of the Lyman-$\alpha$ forest
in the redshift interval $2 \leq z \leq 5$ and $\Delta t_{0} = 10
\; {\rm years}$.

Altogether, encouraged by the prospects of a greatly improved
knowledge of the history of $H(z)$ we present several criteria
that may help telling apart interacting from noninteracting dark
energy models.

Section II presents on phenomenological basis some possible
expressions of the interaction term. Sections III and IV introduce
various criteria to distinguish interacting from noninteracting DE
models. Section V focus on the statefinder parameters and compares
their usefulness with the criteria based on the Hubble history.
Finally, section VI summarizes our findings.

\section{Expressions for the interaction between dark energy and dark matter}
Since the dark components are assumed to interact with one another
also non-gravitationally  but not with baryons the conservation
equations take the form
%%%%%%%%%%%%%%%%%%%%%%%%%%%%%%%%%%%%%%%%%%%%%%%%%%%%%%%%%%%%%%%%%%%%%%%%%%%%%%%%
\begin{eqnarray}
\dot{\rho}_{b}\, &+&\, 3H \rho_{b} = 0 \, , \label{conserv1a} \\
\dot{\rho}_{c}\, &+& \, 3 H \rho_{c} = Q \, , \label{conserv1b} \\
\dot{\rho}_{d}\, &+& \, 3 H (1+w) \rho_{d} = - Q \, ,
\label{conserv1c}
\end{eqnarray}
%%%%%%%%%%%%%%%%%%%%%%%%%%%%%%%%%%%%%%%%%%%%%%%%%%%%%%%%%%%%%%%%%%%%%%%%%%%%%%%%
where $Q$ denotes the interaction term and the relationship
$\rho_{b}+ \rho_{c} = \rho_{m}$ is understood. It is worth
noticing that when $Q > 0$ -as we shall consider throughout- the
coincidence problem is alleviated since $\rho_{c}$ decreases more
slowly with expansion and $\rho_{d}$ more quickly, and the problem
may  even be solved in full \cite{infull}. On the other hand, if
$Q$ were negative, the second law of thermodynamics could be
violated \cite{secondlaw}, and the energy densities of CDM or DE
could become negative at high or low redshifts.

In the absence of a fundamental theory for dark energy the
quantity $Q$ cannot be derived from first principles. However, we
may guess likely expressions for it by noting that $Q$ must be
small \cite{note1} (at least lower than $3H \rho_{m}$) and depend
on the energy densities multiplied by a quantity with units of
inverse of time, a rate. We shall deal with two different
possibilities considered in the literature, namely, $(i)$ that the
said rate is proportional to the Hubble factor, and $(ii)$ that it
is just a constant \cite{gabriela1}.

In case $(i)$,  we have that $Q = Q( H\rho_{c}, \, H\rho_{d})$. By
power law expanding this function and retaining just the first
term one follows $Q \simeq \epsilon_{c}\, H \rho_{c} +
\epsilon_{d}\, H \rho_{d}$. Given the lack of information about
the coupling,  it appears advisable to work with just one
parameter rather than two. Thus, three different choices arise,
namely, $\epsilon_{c} = 0$, $\, \epsilon_{d} = 0$, and
$\epsilon_{c} = \epsilon_{d}$. It is worth mentioning that
expressions of this type can be obtained from the scalar-tensor
theory of gravity developed by Kaloper {\it et al.} \cite{kaloper}
-see e.g. \cite{curbelo,zhang}. In the next section we shall
consider successively three expressions for $Q$ with $0<\epsilon
<< 1$, while keeping $w =$ constant throughout, and look for
diagnostics -based in the history of $H(z)$ that may tell
interacting from noninteracting models.

In case $(ii)$ we have $Q = Q( \rho_{c}, \, \rho_{d})$ and
proceeding as before we write
\begin{equation}
Q = 3\, (\Gamma_{c} \, \rho_{c} \, + \, \Gamma_{d} \, \rho_{d})\,
, \label{QRoy1}
\end{equation}
where the $\Gamma_{i}$ ($i = c, d$) quantities denote constant
rates. This expression was inspired in reheating models
\cite{reheating}, curvaton decay \cite{curvaton}, and decay of DM
into radiation \cite{radiation}. Regrettably, one of the energy
densities ($\rho_{m}$ or $\rho_{d}$) becomes negative either at
early or late times \cite{gabriela1} and, generally, it does not
solve the coincidence problem. Nevertheless, we will briefly
consider it in the section after next.

\section{Interaction term $Q = Q( H\rho_{c}, \, H \rho_{d})$}

We begin by assuming that the interaction term takes the form
%%%%%%%%%%%%%%%%%%%%%%%%%%%%%%%%%%%%%%%%%%%%%%%%%%%%%%%%%%%%%%%%%%%%%%%%%
\begin{equation}
Q = 3\epsilon H \rho_{c} \, , \label{Q1}
\end{equation}
%%%%%%%%%%%%%%%%%%%%%%%%%%%%%%%%%%%%%%%%%%%%%%%%%%%%%%%%%%%%%%%%%%%%%%%%%
where the factor $3$ was introduced for mathematical convenience.

\subsection{Decaying cosmological constant}
One way to explain the present small value of the cosmological
constant is to assume a transfer of energy from the quantum vacuum
to particles and/or radiation -i.e., a decaying cosmological
constant (DCC). Along the years a diversity of phenomenological
models have been proposed, many of which are summarized in Ref.
\cite{overduin}. If the quantum vacuum ($w = -1$) slowly decays
into non-baryonic CDM particles, integration of Eqs.
(\ref{conserv1a})-(\ref{conserv1c}) yields
\begin{eqnarray}
\rho_{b} &=&\rho_{b0}\, x^{3}\, , \nonumber \\
\rho_{c} &=&\rho_{c0}\, x^{3(1-\epsilon)}\, , \nonumber\\
\rho_{\Lambda} &=& \rho_{\Lambda 0} + \left(
\frac{\epsilon}{1-\epsilon}\right)\, \rho_{c0}\,
(x^{3(1-\epsilon)} -1) \, .
\label{intgrlambda}
\end{eqnarray}

This kind of models are not purely phenomenological any more for
they have received some persuasive backings from the theoretical
side \cite{ilya-plb, espana-jcap, wang-meng-cqg, maggiore, neven},
which makes their consideration all the more interesting. In
particular, by means of the renormalization group theory the
authors of \cite{espana-jcap} arrive to a simple expression for
$\epsilon$ ($\nu$ in their notation) in terms of the masses of the
super-heavy sub-planckian particles -see their Eq. (4.10). A
rather similar expression was found by removing the flat
space-time contribution  to the vacuum energy of quantum fields in
homogeneous and isotropic universes by employing a technique
similar to that used in the computation of the
Arnowitt-Deser-Misner mass -see Eq. (39) in Ref. \cite{maggiore}.
Moreover, consistency both with primordial nucleosynthesis and
large scale structure formation leads to the constraint $\mid
\epsilon \mid \ll 1$ \cite{espana-jcap}, \cite{julio-jcap}.
Recently, DCC models were revisited by Basilakos {\it et al.}
\cite{basilakos1} and, for some specific choices of $\Lambda(t)$,
confronted with observational data from supernovae type Ia, baryon
acoustic oscillations, the cosmic microwave background shift
parameter, and the growth rate of galaxy clustering.
%\cite{basilakos2}

As said above, the coincidence problem gets alleviated since
%%%%%%%%%%%%%%%%%%%%%%%%%%%%%%%%%%%%%%%%%%%%%%%%%%%%%%%%%%%%%%%%%%%%%%%%%%%%
\[
\left(\frac{|\dot{r}/r|_{DCC}}{|\dot{r}/r|_{\Lambda
CDM}}\right)_{0} = 1- \epsilon
\left[(\Omega_{c}/\Omega_{\Lambda})_{0} \, + \, r_{0} \right] < 1
\, ,
\]

%%%%%%%%%%%%%%%%%%%%%%%%%%%%%%%%%%%%%%%%%%%%%%%%%%%%%%%%%%%%%%%%%%%%%%%%%%%%
i.e., currently the ratio
$r\equiv(\rho_{b}+\rho_{c})/\rho_{\Lambda}$ varies more slowly
than in the $\Lambda$CDM model. This also holds true for the other
interaction terms considered below.

\noindent Using Friedmann's equation, $ 3 H^{2} = \kappa^{2}
 [\rho_{b}+\rho_{c}\, + \rho_{\Lambda}]$, we can write
\begin{equation}
E^2(x) = \Omega_{b0} \, x^{3} \, + \, \Omega_{c0} \,
x^{3(1-\epsilon)} \, + \, \Omega_{\Lambda 0} \, + \, \left(
\frac{\epsilon}{1-\epsilon}\right) \Omega_{c0}
\left[x^{3(1-\epsilon)} - 1 \right]\, ,
\label{Evlambda}
\end{equation}
where $\Omega_{\Lambda0} = 1-\, \Omega_{b0}\, -\, \Omega_{c0}$.
Obviously, in this case $Om(x) \neq \Omega_{m0}$ with $\Omega_{m0}
= \Omega_{b0}\, + \, \Omega_{c0}$, but rather
\begin{equation}
Om(x) = \Omega_{b0}\, + \, \frac{\Omega_{c0}}{1-\epsilon}\,
\frac{x^{3(1-\epsilon)}-1}{x^{3} -1} \, .
\label{om-dcc}
\end{equation}
Thus, the $Om(x)$ criterion would mistake a DCC model by a
quintessence field (phantom field) if applied at low redshifts
(high redshifts). Similar situations occur if the said criterion
is used on models considered below in this paper.

The first and second derivatives of $E^{2}(x)$ with respect to
$x^{3}$ are \newline $ dE^{2}(x)/dx^{3} = \Omega_{b0}+\Omega_{c0}
\, x^{-3 \epsilon} $, and $d^{2}E^{2}(x)/d(x^{3})^{2} = - \epsilon
\, \Omega_{c0}\, x^{-3(1+\epsilon)}$, respectively. Therefore, for
large redshifts the first derivative tends to $\Omega_{b0}$ from
above. Likewise, $E^{2}(x)$ is concave (i.e., $
d^{2}E^{2}(x)/d(x^{3})^{2} < 0$). By contrast, for noninteracting
dark energy fields with $w =$ constant -see Eq.
(\ref{friedmann1})- one has, $\, dE^{2}(x)/dx^{3} = \Omega_{m0}\,
+\, (1-\Omega_{m0}) \, (1+w)\, x^{3w}$. For large redshifts this
expression tends to $\Omega_{m0}$ from above if DE is of
quintessence type and from below if it is phantom. In view that,
observationally, $\Omega_{b0}$ and $\Omega_{m0}$ are separated by
a non small gap (the latter might be about six or seven times
larger than the former) this may serve discriminate DCC models
from noninteracting quintessence and phantom models. One may think
that we would have arrived to the same result just just by
dividing $E^{2}(x)$ by $x^{3}$; i.e., $E^{2}(x)/x^{3} =
\Omega_{m0} \, + \, (1- \Omega_{m0})\, x^{3w}$. However, this is
not true because of the factor $(1+w)$ in the second term of the
derivative which makes the latter tend to $\Omega_{m0}$ faster,
and it is key to tell whether the DE is quintessence's or
phantom's type.

As a corollary, we may say that the sign of the second derivative,
may discriminate DCC from noninteracting phantom models, the
former being negative and the latter positive,
$d^{2}E^{2}(x)/d(x^{3})^{2} = (1- \Omega_{m0})\, w(1+w)\,
x^{3(w-1)}$; only that this criterion can be used also at smaller
redshift than the preceding one.

Likewise, by measuring $H(z)$ at two  separated redshifts, $z_{1}$
and $z_{2}$ with $z_{1} < z_{2}$, they may be discriminated.
Indeed, for noninteracting phantom models the quantity
$\Delta(x_{1}, x_{2}) := E^{2}(x_{2})\, - \, E^{2}(x_{1})$ is
\begin{equation}
\Delta(x_{1}, x_{2}) =  \Omega_{m0} \, \left[x_{2}^{3}
(1-x_{2}^{3w}) \, - \, x_{1}^{3} (1-x_{1}^{3w})\right]\, + \,
x_{2}^{3(1+w)}\, - \, x_{1}^{3(1+w)}\, , \label{Delta1}
\end{equation}
and it can be negative if the redshifts are sufficiently apart. By
contrast, the corresponding quantity for DCC models,
\begin{equation}
\Delta(x_{1}, x_{2}) = \Omega_{b0} \, (x^{3}_{2} -x^{3}_{1})\, +
\, \Omega_{c0} \, \left(\frac{1}{1-\epsilon}\right) \left[
x^{3(1-\epsilon)}_{2} - x^{3(1-\epsilon)}_{1} \right] \, ,
\label{Delta2}
\end{equation}
will always be positive-definite.

Let us assume that, thanks to extra information, we come to know
that the right model must be a DCC one. Note that to discriminate
which one (i.e., its $\epsilon$ value) neither direct use of last
expression nor of Eq. (\ref{Evlambda}) will prove very useful,
given the comparatively wide uncertainty in the value of
$\Omega_{c0}$. However, the following strategy may work. Consider
two competing DCC models (say A and B) and recall that
$\Omega_{b0}$ is reasonably well determined from primordial
nucleosynthesis ($0.036 \leq \Omega_{b0} \leq 0.047$, see, e.g.
Ref. \cite{olive}) together with the present value of the Hubble
factor. The ratio
%%%%%%%%%%%%%%%%%%%%%%%%%%%%%%%%%%%%%%%%%%%%%%%%%%%%%%%%%%%%%%%%%%%%%%%%%%%%%%%%%%%%%%
\begin{equation}
R_{\Delta}: = \frac{\left[\Delta(x_{1}, x_{2}) \, -\,
\Omega_{b0}\, (x^{3}_{2}-x^{3}_{1})
\right]_{A}}{\left[\Delta(x_{1}, x_{2}) \, -\,  \Omega_{b0}\,
(x^{3}_{2}-x^{3}_{1}) \right]_{B}} =
\left(\frac{1-\epsilon_{B}}{1-\epsilon_{A}}\right)\,
\frac{x^{3(1-\epsilon_{A})}_{2}
-x^{3(1-\epsilon_{A})}_{1}}{x^{3(1-\epsilon_{B})}_{2}
-x^{3(1-\epsilon_{B})}_{1}} \label{ratio1}
\end{equation}
%%%%%%%%%%%%%%%%%%%%%%%%%%%%%%%%%%%%%%%%%%%%%%%%%%%%%%%%%%%%%%%%%%%%%%%%%%%%%%%%%%%%%
depends on $\epsilon_{A}$ and $\epsilon_{B}$ only. This will
enable us to determine which DCC model, A or B, fits better the
$H(z)$ data.

It should be noted that, in this regard, it looks more
advantageous to use this other ratio
\begin{equation}
R_{Om} := \frac{Om(x_{2})-\Omega_{b0}}{Om(x_{1})-\Omega_{b0}} =
\frac{x_{1}^{3} \, - \, 1}{x_{2}^{3} \, - \, 1}\;
\frac{x_{2}^{3(1-\epsilon)} \, - \, 1}{x_{1}^{3(1-\epsilon)} \, -
\, 1}\, ,
\label{om-ratio1}
\end{equation}
since it can constrain $\epsilon$ more easily.

At any rate, aside from this specific instance, the $Om(x)$
function (Eq. (\ref{om})) is not much useful when dealing with
interacting models since, in these models, CDM and baryonic matter
redshift at different rates with expansion whereby the current
fractional densities, $\Omega_{b0}$ and $\Omega_{c0}$, enter the
expression for $E^{2}$ multiplied by different powers of $x$. This
explains the restricted application range of the said function and
we will not resort to it any more in this paper.

\subsubsection{Decaying dark energy field}
For the more general case $w = {\rm constant} \neq -1$, one has
\begin{eqnarray}
\rho_{b} &=&\rho_{b0}\, x^{3}\, , \nonumber \\
\rho_{c} &=&\rho_{c0}\, x^{3(1-\epsilon)}\, , \nonumber\\
\rho_{d} &=& \rho_{d0}\, x^{3(1+w)} + \left(
\frac{\epsilon}{\epsilon \, +\, w}\right)\, \rho_{c0}\, \left[
x^{3(1+w)}- x^{3(1-\epsilon)}\right] \, . \label{intgrde}
\end{eqnarray}
%%%%%%%%%%%%%%%%%%%%%%%%%%%%%%%%%%%%%%%%%%%%%%%%%%%%%%%%%%%%%%%%%%%%%%%%%%%%%%%
Thereby
%%%%%%%%%%%%%%%%%%%%%%%%%%%%%%%%%%%%%%%%%%%%%%%%%%%%%%%%%%%%%%%%%%%%%%%%%%%%%%%
\begin{equation}
E^2(x) = \Omega_{b0} \,x^{3}\,  + \, \Omega_{c0}\,
x^{3(1-\epsilon)} \, + \, \Omega_{d0} \, x^{3(1+w)}\, + \, \left(
\frac{\epsilon}{\epsilon\, +\, w}\right) \Omega_{c0} \left[
x^{3(1+w)}- x^{3(1-\epsilon)}\right]\, , \label{Ede1}
\end{equation}
%%%%%%%%%%%%%%%%%%%%%%%%%%%%%%%%%%%%%%%%%%%%%%%%%%%%%%%%%%%%%%%%%%%%%%%%%%%%%%%
being
%%%%%%%%%%%%%%%%%%%%%%%%%%%%%%%%%%%%%%%%%%%%%%%%%%%%%%%%%%%%%%%%%%%%%%%%%%%%%%%
\begin{eqnarray}
\frac{dE^{2}(x)}{dx^{3}} &=& \Omega_{b0} \, + \, \Omega_{c0}
(1-\epsilon) x^{-3\epsilon}\, + \, \Omega_{d0} (1+w)x^{3w}\,
 \nonumber \\
 &+&\, \left(\frac{\epsilon}{\epsilon + w}\right)
\Omega_{c0} \left[(1+w)x^{3w} \, -\, (1-\epsilon) x^{-3\epsilon}
\right]\, , \label{fderivative}
\end{eqnarray}
%%%%%%%%%%%%%%%%%%%%%%%%%%%%%%%%%%%%%%%%%%%%%%%%%%%%%%%%%%%%%%%%%%%%%%%%%%%%%%%%
and
%%%%%%%%%%%%%%%%%%%%%%%%%%%%%%%%%%%%%%%%%%%%%%%%%%%%%%%%%%%%%%%%%%%%%%%%%%%%%%%%
\begin{eqnarray}
\frac{d^{2}E^{2}(x)}{d(x^{3})^{2}} &=& -(1-\epsilon) \,
\Omega_{c0}\, x^{-3(1+\epsilon)} \, +\, w(1+w) \, \Omega_{d0}\,
x^{3(w-1)}\,  \nonumber \\
&+&  \left(\frac{\epsilon}{\epsilon + w}\right)\Omega_{c0} \left[
w(1+w)x^{3(w-1)}\, +\, \epsilon(1-\epsilon) x^{-3(1+\epsilon)}
\right]  . \label{sderivative}
\end{eqnarray}
%%%%%%%%%%%%%%%%%%%%%%%%%%%%%%%%%%%%%%%%%%%%%%%%%%%%%%%%%%%%%%%%%%%%%%%%%%%%%%%%

In view of the hypothesis $0< \epsilon << 1$ and the fact that
observation reveals that $|1+w| << 1$, the  right hand side of
(\ref{fderivative}) is dominated by its two first terms. Likewise,
the first term dominates the right hand side of
(\ref{sderivative}). Therefore, for not large redshifts (i.e.,
$z<3$) the normalized Hubble function, $E^{2}(x)$, is growing and
concave. In consequence, its shape will not tell a decaying energy
field (with $w \neq -1$) from DCC models or from noninteracting
quintessence or phantom models with constant equation of state.
However, discrimination will likely come at higher redshifts
since, as before, the first derivative may tell interacting from
noninteracting cosmologies depending on whether $dE^{2}(x)/dx^{3}$
tends to $\Omega_{b0}$ or $\Omega_{c0}$ at large redshifts.

\begin{figure}[th]
\includegraphics[width=5.in,angle=0,clip=true]{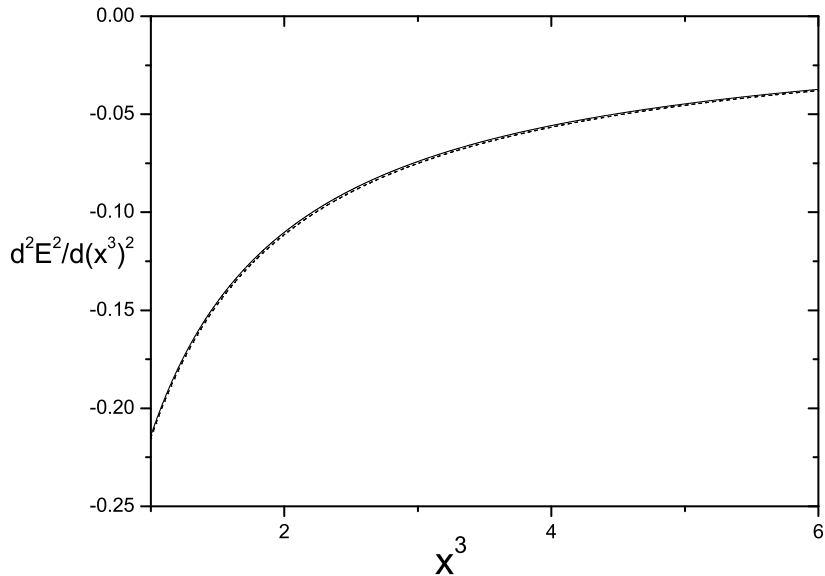}
\caption{Dotted line: plot of the second derivative of $E^{2}$ in
terms of $x^{3}$ for the best fit values, $w = -1.02$, $\epsilon =
0.0006$, with $\Omega_{c0} = 0.230$, and $\Omega_{d0} = 0.725$ of
model III, $Q = 3H \epsilon \rho_{c}$, of Ref.
\cite{he-bin-elcio}. Solid line: same as before but with $\epsilon
= 0$ (no interaction).} \label{fig:sderiv}
\end{figure}

Figure \ref{fig:sderiv} shows the evolution of the second
derivative of $E^{2}(x)$, as given by Eq. (\ref{sderivative}), for
the best fit values of $w$ and $\epsilon$ of model III of  He {\it
et al.} \cite{he-bin-elcio} (dotted line) as constrained by data
from supernovae, baryon acoustic oscillations, cosmic microwave
background, and the value of the Hubble constant, $H_{0}$. Also
shown is the noninteracting case (solid line). The fact that both
lines nearly overlap emphasizes the need for accurate data of
$H(z)$.

%%%%%%%%%%%%%%%%%%%%%%%%%%%%%%%%%%%%%%%%%%%%%%%%%%%%%%%%%%%%%%%%%%%%%%%%%%%%%%
%%%%%%%%%%%%%%%%%%%%%%%%%%%%%%%%%%%%%%%%%%%%%%%%%%%%%%%%%%%%%%%%%%%%%%%%%%%%%%%

In the present case
\begin{eqnarray}
\Delta(x_{1},x_{2}) &=& \Omega_{b0}\, (x^{3}_{2} - x^{3}_{1}) \, +
\, \Omega_{c0}\, \left[x^{3(1-\epsilon)}_{2} -
x^{3(1-\epsilon)}_{1}\right]\, +\, \Omega_{d0}\, \left[
x^{3(1+w)}_{2} -  x^{3(1+w)}_{1}\right] \nonumber \\
&+& \, \left(\frac{\epsilon}{\epsilon+w}\right) \Omega_{c0}\,
\left[x^{3(1+w)}_{2} - x^{3(1+w)}_{1} \ -\, x^{3(1-\epsilon)}_{2}
\,+\, x^{3(1-\epsilon)}_{1}\right] \, . \label{Delta3}
\end{eqnarray}
\noindent For $0< \epsilon <<1$ and $1+w \simeq 0$ the  two last
terms are subdominants, and the ratio defined by the first
equality in (\ref{ratio1}) comes to be
\begin{equation}
R_{\Delta} \simeq \frac{x^{3(1-\epsilon_{A})}_{2}
-x^{3(1-\epsilon_{A})}_{1}}{x^{3(1-\epsilon_{B})}_{2}
-x^{3(1-\epsilon_{B})}_{1}} \; . \label{ratio2}
\end{equation}
Thus, thanks to the factor $(1-\epsilon_{B})/(1-\epsilon_{A})$ on
the right hand side of Eq. (\ref{ratio1}) we hope discriminate DCC
models from interacting dark energy models with $w$ close to the
cosmological constant value, $-1$, and also
discriminate within models featuring the interaction (\ref{Q1}).\\

\subsection{Interaction term proportional to $H \, \rho_{d}$}

We now assume that the interaction term takes the form
%%%%%%%%%%%%%%%%%%%%%%%%%%%%%%%%%%%%%%%%%%%%%%%%%%%%%%%%%%%%%%%%%%%%%%%%%%%%%%%%
\begin{equation}
Q = 3\epsilon H \rho_{d} \,  \label{Q12}
\end{equation}
%%%%%%%%%%%%%%%%%%%%%%%%%%%%%%%%%%%%%%%%%%%%%%%%%%%%%%%%%%%%%%%%%%%%%%%%%%%%%%%%
which may be key to explain the non-vanishing temperature of
sterile neutrinos \cite{hansen}, \cite{jia-zhou}.

Then, the conservation equations
(\ref{conserv1a})-(\ref{conserv1b}) integrate to
%%%%%%%%%%%%%%%%%%%%%%%%%%%%%%%%%%%%%%%%%%%%%%%%%%%%%%%%%%%%%%%%%%%%%%%%%%%%%%%%
\begin{eqnarray}
\rho_{b} &=&\rho_{b0}\, x^{3}\, , \nonumber \\
\rho_{c} &=&\rho_{c0}\, x^{3}+\left(\frac{\epsilon}{w+\epsilon}\right)\, \rho_{d0} \,
\left[1-x^{3(w+\epsilon)}\right]\,x^{3} , \nonumber\\
\rho_{d} &=& \rho_{d0}\, x^{3(1+w+\epsilon)} \, . \label{intgrde2}
\end{eqnarray}
%%%%%%%%%%%%%%%%%%%%%%%%%%%%%%%%%%%%%%%%%%%%%%%%%%%%%%%%%%%%%%%%%%%%%%%%%%%%%%%%
Hence,
%%%%%%%%%%%%%%%%%%%%%%%%%%%%%%%%%%%%%%%%%%%%%%%%%%%%%%%%%%%%%%%%%%%%%%%%%%%%%%%%
\begin{equation}
E^2(x) = x^{3} -\Omega_{d0} \, \left(\frac{w}{w+\epsilon}\right)\,
\left[x^{3} \, - \, x^{3(1+w+\epsilon)}\right] \, .
\label{moresimply}
\end{equation}
%%%%%%%%%%%%%%%%%%%%%%%%%%%%%%%%%%%%%%%%%%%%%%%%%%%%%%%%%%%%%%%%%%%%%%%%%%%%%%%%
 For $w = -1$ (i.e., DCC with interaction given by
Eq. (\ref{Q12})), it reduces to
%%%%%%%%%%%%%%%%%%%%%%%%%%%%%%%%%%%%%%%%%%%%%%%%%%%%%%%%%%%%%%%%%%%%%%%%%%%%%%%%
\begin{equation}
E^{2}(x) = x^{3}\, -\,
\frac{\Omega_{\Lambda0}}{1-\epsilon}\left[x^{3}-x^{3\epsilon}
 \right] \, ,
 \label{Evlambda1}
\end{equation}
%%%%%%%%%%%%%%%%%%%%%%%%%%%%%%%%%%%%%%%%%%%%%%%%%%%%%%%%%%%%%%%%%%%%%%%%%%%%%%%%
which is to be compared with (\ref{Evlambda}). Again, $E^{2}(x)$
is growing and concave whence its graph will tell neither DCC
models with interaction term given by (\ref{Q12}) from DCC models
with interaction term given by (\ref{Q1}) nor these models from
noninteracting quintessence models with $w =$ constant.

Expression (\ref{Evlambda1}) will help constrain $\epsilon$
without need of knowing $\Omega_{\Lambda0}$ since the ratio
%%%%%%%%%%%%%%%%%%%%%%%%%%%%%%%%%%%%%%%%%%%%%%%%%%%%%%%%%%%%%%%%%%%%%%%%%%%%%%%%
\begin{equation}
\frac{[E^{2}(x)-x^{3}]_{x = x_{1}}}{[E^{2}(x)-x^{3}]_{x = x_{2}}}
=  \frac{x_{1}^{3}-x_{1}^{3\epsilon}}{x_{2}^{3}-x_{2}^{3\epsilon}}
\label{ratio4}
\end{equation}
%%%%%%%%%%%%%%%%%%%%%%%%%%%%%%%%%%%%%%%%%%%%%%%%%%%%%%%%%%%%%%%%%%%%%%%%%%%%%%%%
does not depend on that quantity.

Unlike the two previous cases, as inspection of the right hand
side of Eq. (\ref{moresimply}) shows,  the behavior of the
derivative $dE^{2}(x)/dx^{3}$ at high redshifts will not
discriminate interacting from noninteracting cosmological models.
For the discrimination to be possible $\epsilon$ should take
unrealistic high values.

By deriving twice (\ref{moresimply}) we get
%%%%%%%%%%%%%%%%%%%%%%%%%%%%%%%%%%%%%%%%%%%%%%%%%%%%%%%%%%%%%%%%%%%%%%%%%%%%%%%%
\begin{equation}
\frac{d^{2}E^{2}(x)}{d(x^{3})^{2}} = \Omega_{d0} \,
w(1+w+\epsilon)\, x^{3(w+\epsilon-1)}. \label{sderivative2}
\end{equation}
%%%%%%%%%%%%%%%%%%%%%%%%%%%%%%%%%%%%%%%%%%%%%%%%%%%%%%%%%%%%%%%%%%%%%%%%%%%%%%%%

When this latter quantity gets accurately determined, we will be
able to set useful constraints on $w$ and $\epsilon$ since its
ratio for two decaying dark energy (DDE) models (say $A$ and $B$)
does not depend on the $\Omega$ parameters,
%%%%%%%%%%%%%%%%%%%%%%%%%%%%%%%%%%%%%%%%%%%%%%%%%%%%%%%%%%%%%%%%%%%%%%%%%%%%%%%%
\begin{equation}
\frac{d^{2}E^{2}/d(x^{3})^{2}|_{A}}{d^{2}E^{2}/d(x^{3})^{2}|_{B}}
=\frac{w_{A}(1+w_{A}+\epsilon_{A})}{w_{B}(1+w_{B}+\epsilon_{B})}\,
\frac{x^{3(w_{A}+\epsilon_{A})}}{x^{3(w_{B}+\epsilon_{B})}}\,.
\label{ratio3}
\end{equation}
%%%%%%%%%%%%%%%%%%%%%%%%%%%%%%%%%%%%%%%%%%%%%%%%%%%%%%%%%%%%%%%%%%%%%%%%%%%%%%%%

Once again $E^{2}(x)$, given by (\ref{moresimply}), is a growing
function of redshift, concave for quintessence fields and convex
for phantom fields. Accordingly, the shape of $E^{2}(x)$ can tell
decaying quintessence fields from decaying phantom fields but not
from noninteracting DE fields.

\begin{figure}[th]
\includegraphics[width=5.in,angle=0,clip=true]{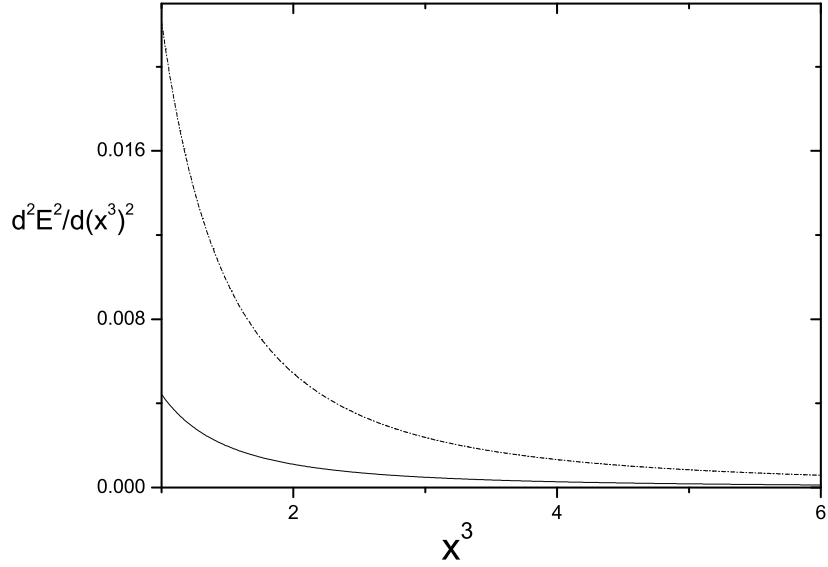}
\caption{Dotted line: plot of the second derivative of $E^{2}$ in
terms of $x^{3}$ for the best fit values, $w = -1.03$, $\epsilon =
0.024$, with  $\Omega_{d0} = 0.717$ of model II, $Q = 3H \epsilon
\rho_{d}$, of Ref. \cite{he-bin-elcio}. Solid line: same as before
but with $\epsilon = 0$ (no interaction).} \label{fig:sderiv2}
\end{figure}

Figure \ref{fig:sderiv2} depicts the evolution of the second
derivative of $E^{2}(x)$, as given by Eq. (\ref{sderivative2}),
for the best fit values of $w$ and $\epsilon$ of model II of Ref.
\cite{he-bin-elcio} (dotted line) as constrained by observational
data. The solid line corresponds to the noninteracting case (solid
line). As it is apparent the lower the redshift, the easier the
discrimination.
%%%%%%%%%%%%%%%%%%%%%%%%%%%%%%%%%%%%%%%%%%%%%%%%%%%%%%%%%%%%%%%%%%%%%%%%%%%%%%%
%%%%%%%%%%%%%%%%%%%%%%%%%%%%%%%%%%%%%%%%%%%%%%%%%%%%%%%%%%%%%%%%%%%%%%%%%%%%%%%%

\subsection{Interaction term proportional to $H \, (\rho_{c}+\rho_d)$}

Assuming the interaction term
%%%%%%%%%%%%%%%%%%%%%%%%%%%%%%%%%%%%%%%%%%%%%%%%%%%%%%%%%%%%%%%%%%%%%%%%%%%%%%%%%%
\begin{equation}
Q = 3\epsilon H (\rho_{c}+\rho_{d})\, , \label{Q13}
\end{equation}
%%%%%%%%%%%%%%%%%%%%%%%%%%%%%%%%%%%%%%%%%%%%%%%%%%%%%%%%%%%%%%%%%%%%%%%%%%%%%%%%%%
widely considered in the literature -see, e.g.
\cite{infull,Q3,gabriela1}-, equations
(\ref{conserv1a})-(\ref{conserv1c}) integrate to
%%%%%%%%%%%%%%%%%%%%%%%%%%%%%%%%%%%%%%%%%%%%%%%%%%%%%%%%%%%%%%%%%%%%%%%%%%%%%%%%%%
\begin{eqnarray}
\rho_{b} &=&\rho_{b0}\, x^{3}\, , \nonumber \\
\rho_{c} &=&C_1\,x^{\gamma_1}+C_2\,x^{\gamma_2} , \nonumber\\
\rho_{d} &=&
\frac{1}{2\,\epsilon}\,\left[-C_1(A+B)+C_2\,(B-A)\,x^{-3\,B}\right]\,
x^{\gamma_1} \, , \label{intgrde2}
\end{eqnarray}
%%%%%%%%%%%%%%%%%%%%%%%%%%%%%%%%%%%%%%%%%%%%%%%%%%%%%%%%%%%%%%%%%%%%%%%%%%%%%%%%%%
respectively.

\begin{figure}[th]
\includegraphics[width=3.in,angle=0,clip=true]{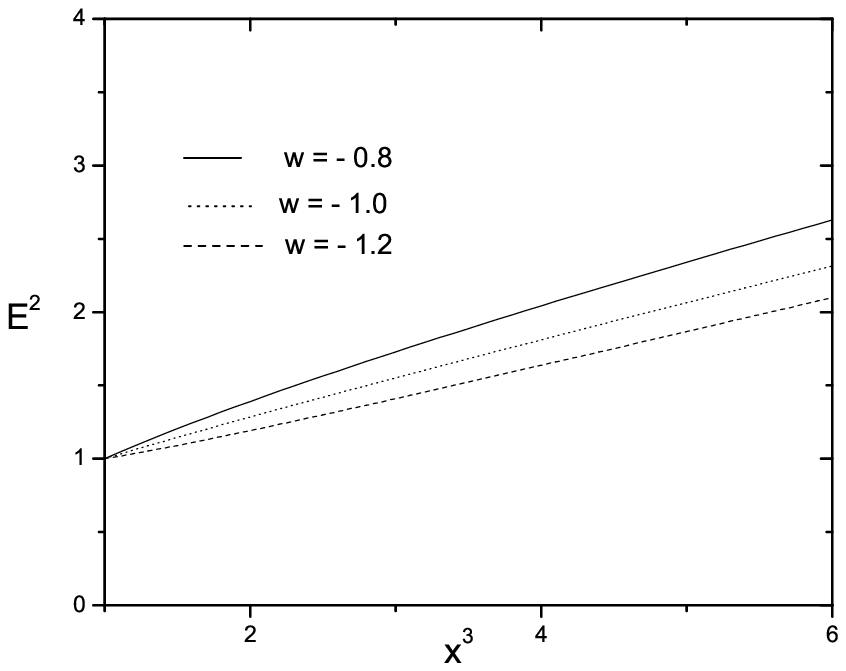}
\includegraphics[width=3.2in,angle=0,clip=true]{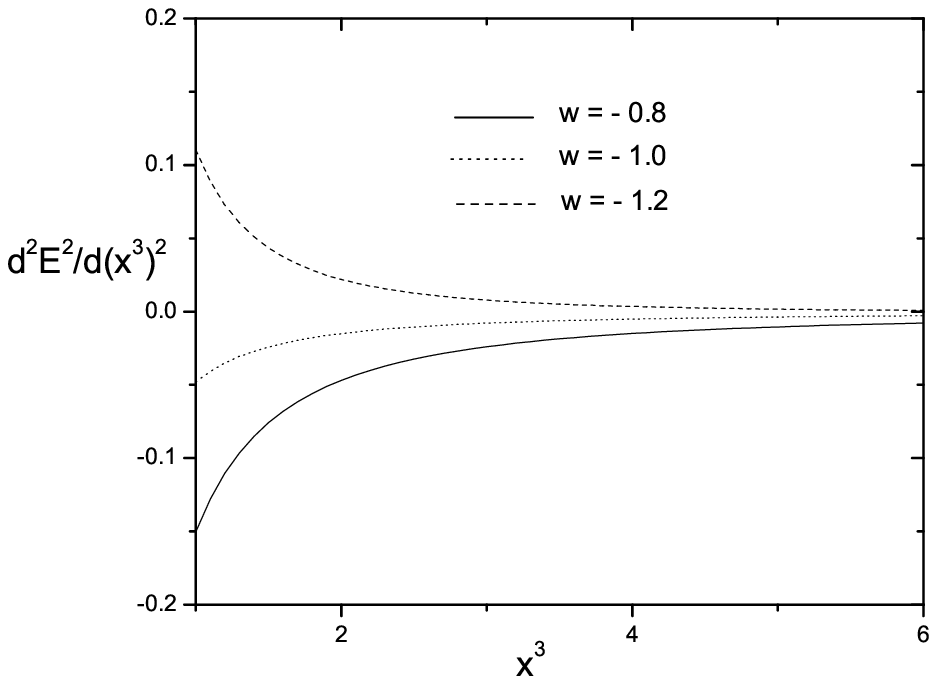}
\caption{Left panel: Graphs of $E^{2}$ vs. $x^{3}$ as given by Eq.
(\ref{Ede13}) for three values of $w$. Right panel: Graphs of the
second derivative of $E^{2}$ with respect to $x^{3}$  for the same
three values of $w$. Notice that only phantom models present
convex curves for reasonable values of the parameters. In drawing
the graphs we took $\Omega_{b0} = 0.04$, $\Omega_{c0} = 0.26$,
$\Omega_{d0} = 0.70$, and $\epsilon = 0.05$.} \label{figure1}
\end{figure}

\noindent Here
%%%%%%%%%%%%%%%%%%%%%%%%%%%%%%%%%%%%%%%%%%%%%%%%%%%%%%%%%%%%%%%%%%%%%%%%%%%%%%%%%%
\[
\gamma_1=\frac{3}{2}\,[2+w+\sqrt{w(w+4\epsilon)}]\,,\,\;\;\;\;\gamma_2=\frac{3}{2}\,[2+w-\sqrt{w(w+4\epsilon)}]\,,
\]
%%%%%%%%%%%%%%%%%%%%%%%%%%%%%%%%%%%%%%%%%%%%%%%%%%%%%%%%%%%%%%%%%%%%%%%%%%%%%%%%%%
\[
C_1=\frac{1}{2\,B}\left[(B-A)\,\rho_{c0}-2\,\epsilon\,\rho_{d0}\right]\,,\;\,\;\,
C_2=\frac{1}{2\,B}\left[(A+B)\,\rho_{c0}+2\,\epsilon\,\rho_{d0}\right],
\]
%%%%%%%%%%%%%%%%%%%%%%%%%%%%%%%%%%%%%%%%%%%%%%%%%%%%%%%%%%%%%%%%%%%%%%%%%%%%%%%%%%
\[
A=(w+2\,\epsilon)\,,\;\;\;\;\mbox{and}
\,\,\;\;\;\;B=\sqrt{w(w+4\epsilon)}.
\]
%%%%%%%%%%%%%%%%%%%%%%%%%%%%%%%%%%%%%%%%%%%%%%%%%%%%%%%%%%%%%%%%%%%%%%%%%%%%%%%%%%
\noindent As in the two previous cases, the three energy densities
remain semi-positive definite for all $x$. Therefore,
%%%%%%%%%%%%%%%%%%%%%%%%%%%%%%%%%%%%%%%%%%%%%%%%%%%%%%%%%%%%%%%%%%%%%%%%%%%%%%%%%%
\begin{equation}
E^2(x) = \Omega_{b0} \,x^{3}\,  +
\frac{1}{2\,\epsilon}\,\left\{\Omega_{C_1} [2\epsilon- (A+B)] \,
-\Omega_{C_2}\,(A-B) \, x^{-3B} \right\}\, x^{\gamma_1}\, +
\Omega_{C_2} \, x^{\gamma_{2}} \, , \label{Ede13}
\end{equation}
where
%%%%%%%%%%%%%%%%%%%%%%%%%%%%%%%%%%%%%%%%%%%%%%%%%%%%%%%%%%%%%%%%%%%%%%%%%%%%%%%%%%
\[
\Omega_{C_1}=\frac{1}{2\,B}\left[\Omega_{c0}\,(B-A)-2\,\epsilon\,\Omega_{d0}\right]\,,
\;\,\;\,\mbox{and}\;\;\;\;\;\;\;
\Omega_{C_2}=\frac{1}{2\,B}\left[\Omega_{c0}\,(A+B)+2\,\epsilon\,\Omega_{d0}\right]\,
,
\]
%%%%%%%%%%%%%%%%%%%%%%%%%%%%%%%%%%%%%%%%%%%%%%%%%%%%%%%%%%%%%%%%%%%%%%%%%%%%%%%%%%
also stays non-negative as it should. Further, $E^{2}$ is a
growing function of $x^{3}$, concave for quintessence and convex
for phantom -see Fig. \ref{figure1}- which will help discriminate
phantom from quintessence DE in interacting models with $Q$ given
by Eq. (\ref{Q13}). However, as  is apparent, for $x^{3} \geq 5$
the graphs show a nearly straight line behavior whereby one must
focus on redshifts below, say, $1.7$.

Inspection of the right hand side of (\ref{Ede13}) shows that for
interacting models ($\epsilon> 0$) the derivative
$dE^{2}(x)/dx^{3}$ tends to $\Omega_{b0}$ at high redshifts.
Accordingly if observation reveals that behavior, it would be
suggestive of interaction.

%%%%%%%%%%%%%%%%%%%%%%%%%%%%%%%%%%%%%%%%%%%%%%%%%%%%%%%%%%%%%%%%%%%%%%%%%%%%%
%%%%%%%%%%%%%%%%%%%%%%%%%%%%%%%%%%%%%%%%%%%%%%%%%%%%%%%%%%%%%%%%%%%%%%%%%%%%%%%

\begin{figure}[th]
\includegraphics[width=5in,angle=0,clip=true]{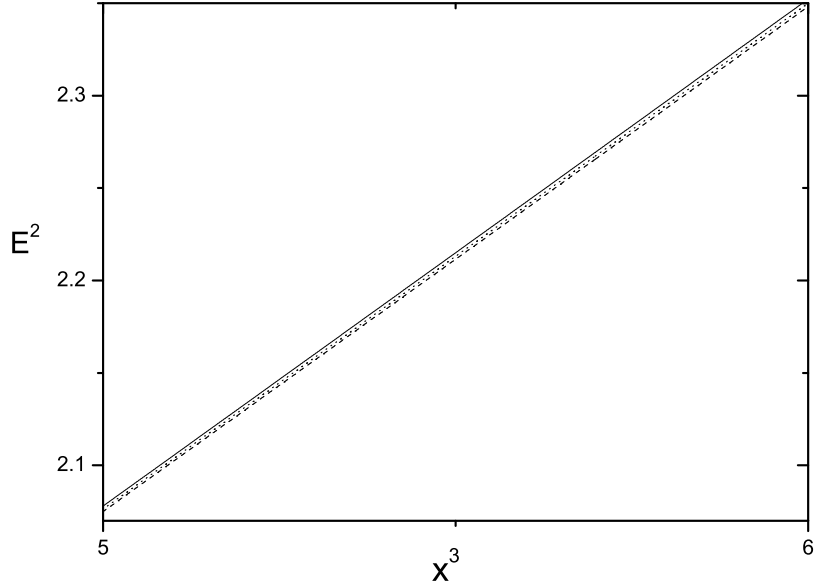}
\caption{Dotted line: plot of $E^{2}$ in terms of $x^{3}$ for the
best fit values, $w = -1.03$, $\Omega_{c0} = 0.230$, and
$\Omega_{d0} = 0.725$, $\epsilon = 0.0006$, (dotted line) and the
$\epsilon$ values corresponding to the $1\sigma$ errors, $0.0011$
(dashed) and $\simeq 0$ (solid) of model IV, $Q = 3H \epsilon
(\rho_{c} \, + \, \rho_{d})$, of Ref. \cite{he-bin-elcio}. We just
depict the $5 \leq x^{3} \leq 6$ interval because the difference
between the interacting and noninteracting case augments with
redshift.} \label{fig:e2}
\end{figure}

\begin{figure}[th]
\includegraphics[width=5in,angle=0,clip=true]{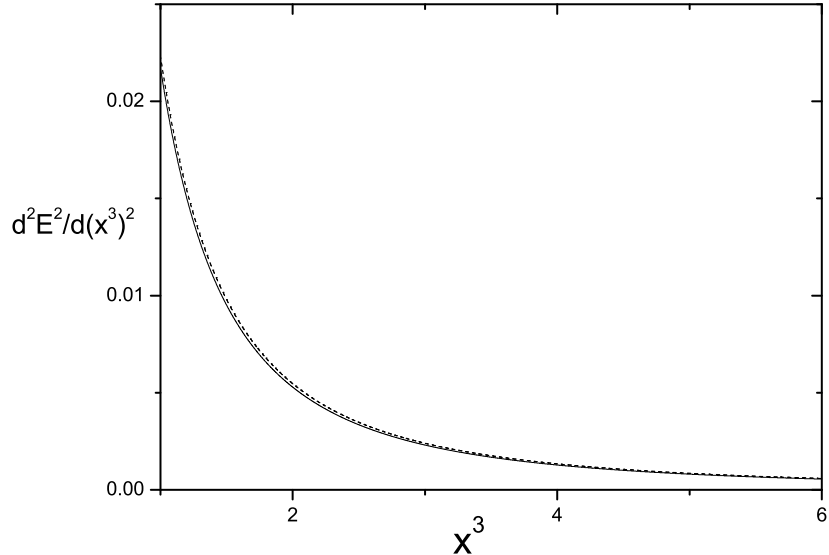}
\caption{Dotted line: plot of the second derivative of $E^{2}$
for the best fit values, $w = -1.03$, $\epsilon = 0.0006$, with
$\Omega_{c0} = 0.230$, and $\Omega_{d0} = 0.725$, of model IV, $Q
= 3H \epsilon (\rho_{c} \, + \, \rho_{d})$, of Ref.
\cite{he-bin-elcio}. Solid line: same as before but with $\epsilon
\simeq 0$ (no interaction).} \label{fig:sderiv3}
\end{figure}

Figure \ref{fig:e2} shows the evolution of $E^{2}$ for the best
fit value of model IV ($\epsilon =0.0006$) of \cite{he-bin-elcio}
(dotted line) and the values corresponding to the $1\sigma$ error,
$\epsilon = 0.0011$ (dashed) and $\epsilon \simeq 0$ (solid);
obviously the latter practically coincides with the noninteracting
case. Likewise, Fig. \ref{fig:sderiv3} shows the evolution of
$d^{2}E^{2}(x)/d(x^{3})^{2}$, for the best fit values of the same
model (dotted line) and for the noninteracting case (solid line).
As is apparent the graphs practically overlap.
%%%%%%%%%%%%%%%%%%%%%%%%%%%%%%%%%%%%%%%%%%%%%%%%%%%%%%%%%%%%%%%%%%%%%%%%%%%%%%%%%%%%
%%%%%%%%%%%%%%%%%%%%%%%%%%%%%%%%%%%%%%%%%%%%%%%%%%%%%%%%%%%%%%%%%%%%%%%%%%%%%%%%%%%%
%%%%%%%%%%%%%%%%%%%%%%%%%%%%%%%%%%%%%%%%%%%%%%%%%%%%%%%%%%%%%%%%%%%%%%%%%%%%%%%%%%%%
%%%%%%%%%%%%%%%%%%%%%%%%%%%%%%%%%%%%%%%%%%%%%%%%%%%%%%%%%%%%%%%%%%%%%%%%%%%%%%%%%%%%

\section{Interaction terms of the type $Q = Q(\rho_{c}, \, \rho_{d})$}

In case $(ii)$ -i.e., for $Q$ given by Eq. (\ref{QRoy1})-
analytical solution can be found, at most, for one of the
conservation equations, either (\ref{conserv1b}) or
(\ref{conserv1c}), and this only if $\Gamma_{d} = 0$, or
$\Gamma_{c} = 0$. Hence, at least one these equations must be
solved numerically whereby no analytical expression exists for the
Hubble factor. This limits the usefulness of $ H(z)$ to set
diagnostics on the nature of DE.

Let us consider first the case  $\Gamma_{c} = 0$ and $\Gamma_{d} >
0$. In this instance, the currently phase of DE domination would
be a transient one, the smaller $\Gamma_{x}$, the longer the DE
domination.

\begin{figure}[th]
\includegraphics[width=3.in,angle=0,clip=true]{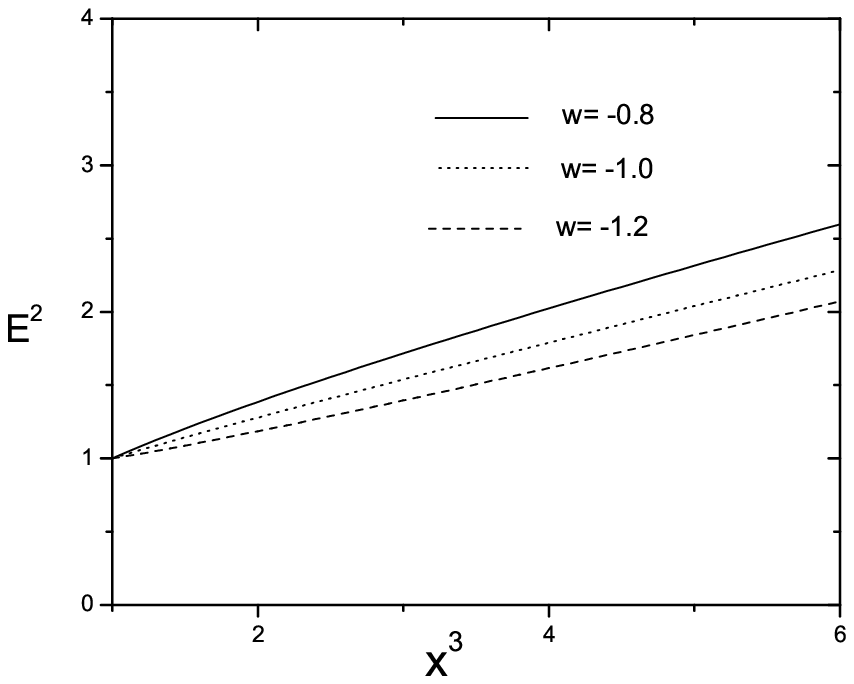}
\includegraphics[width=3.2in,angle=0,clip=true]{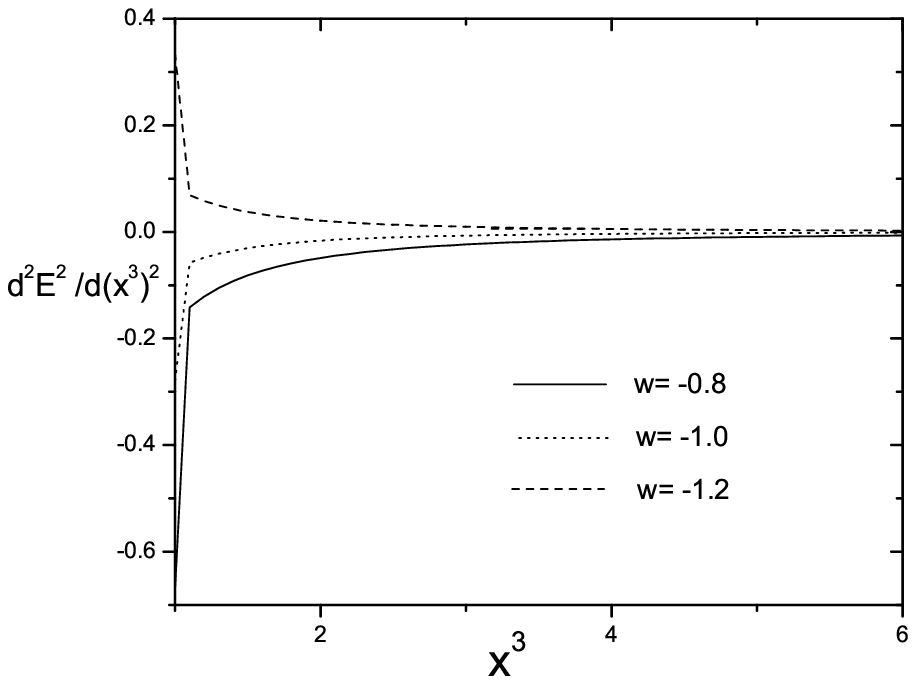}
\caption{Left panel: Graphs of $E^{2}$ vs. $x^{3}$ assuming $Q = 3
\, \Gamma_{d} \, \rho_{d}$ with $\Gamma_{d} = 0.1 \, H_{0}$ for
three values of $w$. Right panel: Graphs of the second derivative
of $E^{2}$ with respect to $x^{3}$ for the same three values of
$w$. In drawing the graphs we took $\Omega_{b0} = 0.04$,
$\Omega_{c0} = 0.26$, and $\Omega_{d0} = 0.70$.} \label{figure2}
\end{figure}

Left and right panels of panel of Fig. \ref{figure2} depict the
evolution of $E^{2}$ and its second derivative in terms of $x^{3}$
for $\Gamma_{d} = 0.1 \, H_{0}$.

%%%%%%%%%%%%%%%%%%%%%%%%%%%%%%%%%%%%%%%%%%%%%%%%%%%%%%%%%%%%%%%%%%%%%%%%%
%%%%%%%%%%%%%%%%%%%%%%%%%%%%%%%%%%%%%%%%%%%%%%%%%%%%%%%%%%%%%%%%%%%%%%%%%%

As it can be seen, $d^{2}E^{2}(x)/d(x^{3})^2$ presents a very fast
increase (decrease) for phantom and DCC (for quintessence) at
small redshifts ($z \ll 1)$. This feature is absent in
noninteracting models as well as in models whose interaction term
is of the general type considered in the previous section.

%%%%%%%%%%%%%%%%%%%%%%%%%%%%%%%%%%%%%%%%%%%%%%%%%%%%%%%%%%%%%%%%%%%%%%%%%%%%%
\begin{figure}[th]
\includegraphics[width=3.in,angle=0,clip=true]{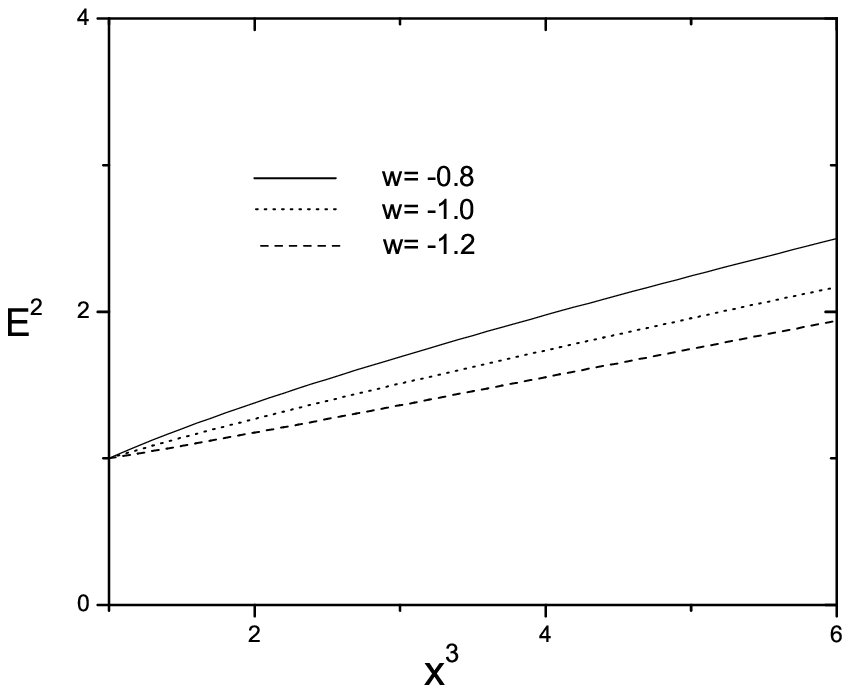}
\includegraphics[width=3.2in,angle=0,clip=true]{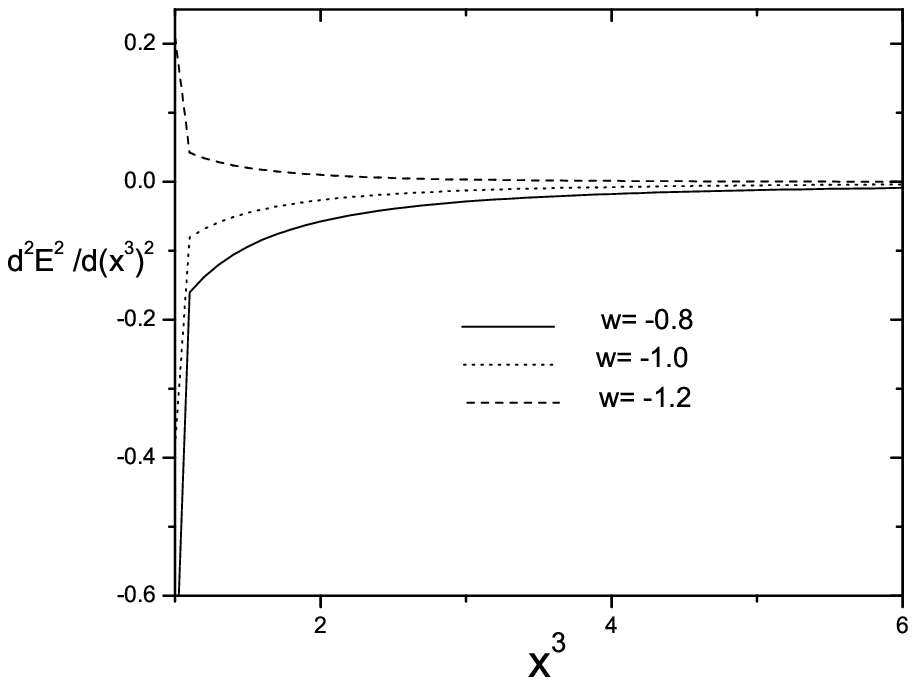}
\caption{Left panel: Graphs of $E^{2}$ vs. $x^{3}$ assuming $Q = 3
\, (\Gamma_{c} \, \rho_{c} \, + \, \Gamma_{d} \, \rho_{d})$ with
$\Gamma_{c} = \Gamma_{d} = 0.1 \, H_{0}$ for three values of $w$.
Right panel: Graphs of the second derivative of $E^{2}$ with
respect to $x^{3}$ for the same three values of $w$. In drawing
the graphs we took $\Omega_{b0} = 0.04$, $\Omega_{c0} = 0.26$,
$\Omega_{d0} = 0.70$.} \label{figure3}
\end{figure}
%%%%%%%%%%%%%%%%%%%%%%%%%%%%%%%%%%%%%%%%%%%%%%%%%%%%%%%%%%%%%%%%%%%%%%%%%%%%%%

As a second example, let us consider the interaction term  given
by Eq. (\ref{QRoy1}) with $\Gamma_{c} = \Gamma_{d} = 0. 1 \,
H_{0}$. As can be seen in Fig. \ref{figure3}, the behavior of the
second derivative, $d^{2}E^{2}(x)/d(x^{3})^2$, at small redshifts
is very similar to the previous case whence we cannot tell one
interaction from the other. Accordingly, an initial steep slope
followed by an abrupt change to a much milder slope appears to be
a general characteristic of interacting models in which $Q$ is
proportional to the densities of DM or DE or to a combination of
them. We hope this will serve to discriminate this set of
interactions from the rest.

\section{The statefinder parameters}
 Years ago Sahni {\it et al.} \cite{statefinders1} and Alam
 {\it et al.} \cite{statefinders2}-based on the dimensionless,
 geometrical pair of parameters $R$ and $S$- gave a somewhat
 similar diagnostic of dark energy. These are defined
in terms of the third temporal derivative of the scale
factor, namely, $R =\stackrel{...}{a}/(aH^{3})$ and $S = (R -1)/(3
\left(q - \textstyle{1\over{2}}\right))$, where $q$ stands for the
deceleration parameter. The usefulness of the said pair stems from
the fact that different models are characterized by different
graphs in the $(S,R)$ plane; in particular, the $\Lambda$CDM model
is associated to the single point $(0,1)$. So, they may help
elucidate the right DE candidate.

When these parameters are calculated for cosmological models in
which DE and DM interact with each other it is seen that the
interaction first appears in the third derivative of the scale
factor \cite{statefinders3}. In particular, for $Q = 3 \epsilon\,
H \, \rho_{c}$ one has
%%%%%%%%%%%%%%%%%%%%%%%%%%%%%%%%%%%%%%%%%%%%%%%%%%%%%%%%%%%%%%%%%%%%%%%%%%%%%%%%%%
\begin{equation}
R = 1 \, + \, \frac{9}{2}\, w (1 \, + \, w \,\Omega_{d})\;
\frac{(1+w)\Omega_{d} \, + \, \epsilon\, \Omega_{c}}{\Omega_{b} \,
+ \, \Omega_{c} \, + \, (1+w)\, \Omega_{d}} \, .
\label{statefinder1}
\end{equation}
%%%%%%%%%%%%%%%%%%%%%%%%%%%%%%%%%%%%%%%%%%%%%%%%%%%%%%%%%%%%%%%%%%%%%%%%%%%%%%%%%%

More specifically if one wished to resort to the present value of
$R$ to discriminate the $\Lambda$CDM model from the DCC model ($w
= -1$, $Q = 3 \epsilon H \, \rho_{c}$) of sub-section III.A, one
would find $R_{0} = 1\, - \, \textstyle{9 \over{2}} \, \epsilon \,
\Omega_{c0}$.

If, instead, the interaction is governed by Eq. (\ref{QRoy1}) one
obtains
%%%%%%%%%%%%%%%%%%%%%%%%%%%%%%%%%%%%%%%%%%%%%%%%%%%%%%%%%%%%%%%%%%%%%%%%%%%%%%%%%%
\begin{equation}
R = 1 \, + \, \frac{9}{2}\, w (1 \, + \, w \,\Omega_{d})\;
\frac{(1+w)\Omega_{d} \, + \, \frac{1}{H}(\Gamma_{c} \,
\Omega_{c}\, + \, \Gamma_{d} \, \Omega_{d})}{\Omega_{b} \, + \,
\Omega_{c} \, + \, (1+w)\, \Omega_{d}} \, . \label{statefinderRoy}
\end{equation}
%%%%%%%%%%%%%%%%%%%%%%%%%%%%%%%%%%%%%%%%%%%%%%%%%%%%%%%%%%%%%%%%%%%%%%%%%%%%%%%%%%

This suggests that once the $\Omega$ parameters come to be
accurately  determined at different redshifts, the statefinder
parameters will not only tell whether the interaction really
exists but also will help discern the expression for $Q$. However,
for the time being, the wide uncertainties about these seriously
restrict the utility of the $R,S$ pair in this respect.

Altogether, the diagnostics founded on the $E^{2}$ and
$dE^{2}/dx^{3}$ functions appear more advantageous since the
interaction parameters (i.e., the $\epsilon$ and $\Gamma_{i}$
quantities), enter the Hubble factor upon integration (analytical
or numerical) of the conservation equations
(\ref{conserv1a})-(\ref{conserv1c}).

\section{Discussion}
Interacting models of dark energy are well motivated,
substantially alleviate the coincidence problem, and show
compatibility with observation. Thus the question arises, ``can
they be discriminated from noninteracting models?" For the time
being, the answer is not given our still imperfect knowledge of
basic cosmological parameters, such as the Hubble factor. The
latter is key in constraining models with observational data from
baryon acoustic oscillations, the R-shift parameter of cosmic
background radiation, the matter growth parameter, etc.
Nevertheless, the situation is to improve greatly in the not far
future thanks to a variety of upcoming and planed experiments.

To extract the invaluable information about the nature of dark
energy encrypted in the history of the Hubble factor is does not
suffice to have abundant, accurate data of $H(z)$. One must first
set up simple and practical criteria to be used once these data
become available. We proposed several criteria, of notable
mathematical simplicity, based on an accurate knowledge of the
said history -something we reasonably hope to have at our disposal
not very soon but, at least, in the foreseeable future.

Specifically if $d^{2}E^{2}/d(x^{3})^{2}$ results positive
(negative), then the dark energy is of phantom type (either
quintessence or a decaying cosmological constant). Only if the DE
is not-decaying cosmological constant, the graph of $E^{2}$ vs.
$x^{3}$ will yield a straight line.

To discern whether DE is interacting (with $Q = Q(H \rho_{c}, \, H
\rho_{d})$) or not the behavior of $d E^{2}(x)/dx^{3}$ at high
redshifts must be studied. If this derivative tends to the present
value of the fractional baryon density, $\Omega_{b0}$, then there
will be grounds to believe that is it interacting with the
interaction term $Q$ given either by Eq. (\ref{Q1}) or
(\ref{Q13}). If it tends to $\Omega_{m0}$, then we may strongly
suspect that it is not interacting; and, in particular, if it
coincides with $\Omega_{m0}$ independently of redshift, then it
will most likely be a conserved cosmological constant -see Eq.
(\ref{friedmann1}). Finally, if the redshift function
$dE^{2}(x)/dx^{3}$ does not tend to any of these two quantities,
we may conclude that either DE is interacting with $Q$ given by
Eq. (\ref{Q12}) or some other law not considered here, or that in
reality $w$ is not a constant but a function of redshift.
Obviously, this would complicate matters very much because it
would call for the introduction of further parameters in the
analysis -something that we defer to a future work.

If the plot of $d^{2}E^{2}(x)/d(x^{3})^{2}$ vs. $x^{3}$ exhibits a
very steep slope at redshifts much smaller than unity, we may
conclude that the interaction obeys Eq. (\ref{QRoy1}) (with
$\Gamma_{c}$ and $\Gamma_{d}$ non-negative). Its absence excludes
such interaction, but it is very hard to specify whether the slope
is steep enough. On the other hand, since this criterion deals
with the slope of the second derivative extremely accurate data of
$ H(z)$ at these redshifts must be employed.

The statefinder parameters
\cite{statefinders1,statefinders2,statefinders3} will be very
useful to rule out cosmological models, and tell apart different
interacting models, but first the histories of $\Omega_{c}(z)$ and
$\Omega_{d}(z)$ must be accurately established in a
model-independent manner. This looks more challenging than
obtaining an accurate history of the Hubble factor.

Since the interaction alters the abundance of dark matter in the
past two additional ways to ascertain if the DE is interacting or
not can be implemented, namely, to examine the behavior of the
growth factor, and to study  the weak lensing power spectrum. If
-as we assumed throughout- the DE decays into dark matter, then it
must have been less dark matter in the past than in the
corresponding noninteracting model with the same $w$ and,
consequently, the growth factor must be smaller. On the other
hand, this would imply a decline of the lensing spectrum (the
opposite would hold true if DM decayed into DE). Both effects have
been considered by the authors of Ref. \cite{gabriela2}. However,
they do not take $w$ constant but assume the
Chevalier-Polarski-Linder ansatz \cite{ch-p-l} with very specific
values for the parameters entering that expression. This is why
comparison with our work does not seem straightforward. At any
rate, both studies may be seen as complementary.

While the criteria introduced in this paper should be helpful in
deciphering the nature of DE they should be used alongside the
analysis and interpretation of data coming from a variety of
sources: supernovae type Ia, cosmic microwave background radiation
(CMBR), matter power spectrum, baryon acoustic oscillations, weak
lensing, and evolution of galaxy clusters. The latter lies at the
core of recent studies about the dynamical equilibrium of the
clusters which seem to favor a decay of DE into DM over other
possible scenarios \cite{bwang}.

In particular, a promising way to test the existence of the
interaction is to consider its impact on the integrated
Sachs-Wolfe effect. Since the former alters the evolution rate of
gravitational potentials it must affect the frequency shift
experienced by CMBR photons crossing collapsing structures. Again,
studies of this kind hint at the existence of the interaction with
DE decaying into DM \cite{gmo-isw,jiang-hua}.

We restricted ourselves to constant $w$ and vanishing spatial
curvature. While this automatically limits the scope of our work
we feel it does not do it seriously, at least no more than
otherwise since it would introduce additional unknown parameters
that would compromise the said scope in other ways.

\acknowledgments{Thanks are due to the anonymous refreee and
Fernando Atrio-Barandela for constructive comments and criticisms
on a earlier draft of this paper, and Eric Linder and Craig Hogan
for drawing our attention to Refs. \cite{schlegel} and
\cite{craig1}, respectively. This work was supported from
``Comisi\'{o}n Nacional de Ciencias y Tecnolog\'{\i}a" (Chile)
through the FONDECYT Grant No. 1070306 (SdC) and  No. 1090613
(RH). D.P. acknowledges ``FONDECYT-Concurso incentivo a la
cooperi\'{o}n internacional" No. 7100025, and is grateful to the
``Instituto de F\'{\i}sica" for warm hospitality; also D.P.
research was partially supported by the ``Ministerio Espa\~{n}ol
de Educaci\'{o}n y Ciencia" under Grant No. FIS2006-12296-C02-01.}


\begin{thebibliography}{99}
\bibitem{note1}
If it were large and positive, DM would dominate the expansion
today. On the other hand, if $Q$ were large and negative, the
Universe would have been dominated by DE practically from the
outset and galaxies would not have come into existence.
\end{thebibliography}
\end{document}